**New pure shear acoustic surface waves guided by cuts in magneto-electro-elastic materials**

Arman Melkumyan

*Department of Mechanics, Yerevan State University, Alex Manoogyan Str. 1, Yerevan 375025, Armenia*

E-mail: melk_arman@yahoo.com

**ABSTRACT** It is shown that new pure shear acoustic surface waves with five different velocities can be guided by stress free plane cuts with different magneto-electrical properties in magneto-electro-elastic materials. The possibility for the surface waves to be guided by a cut in pairs, which is reported in this paper, is new in magneto-electro-elastic materials and has no counterpart in piezoelectric materials. The five velocities of propagation of the surface waves are obtained in explicit forms. It is shown that the possibility for the surface waves to be guided in pairs disappears and the number of surface waves decreases from 5 to 1 if the magneto-electro-elastic material is changed to a piezoelectric material.

Bleustein [1] and Gulyaev [2] have shown that an elastic shear surface wave can be guided by the free surface of piezoelectric materials in class 6 mm, and later Danicki [3] has described the propagation of a shear surface wave guided by an embedded conducting plane. In this paper the existence of pure shear acoustic surface waves guided by infinite plane cuts in transversely isotropic magneto-electro-elastic [4-8] materials in class 6 mm is investigated. The medium can also be considered as a system of two magneto-electro-elastic half-spaces with infinitesimally thin layers on their traction free plane surfaces with different magneto-electric properties. Discussing different magneto-electrical boundary conditions on the free surfaces, pure shear surface waves with 5 different velocities of propagation are obtained. In this paper it is also reported that in magneto-electro-elastic materials the surface waves can be guided in pairs, which has no counterpart in piezoelectric as well as in elastic materials. It is expected that these waves will have numerous applications in surface acoustic wave devices.

Let $x_1$, $x_2$, $x_3$ denote rectangular Cartesian coordinates with $x_3$ oriented in the direction of the sixfold axis of a magneto-electro-elastic material in class 6 mm. By introducing electric potential $\varphi$ and magnetic potential $\phi$, so that $E_1 = -\varphi_{,1}$, $E_2 = -\varphi_{,2}$, $H_1 = -\phi_{,1}$, $H_2 = -\phi_{,2}$, the five partial differential equations which govern the mechanical displacements $u_1$, $u_2$, $u_3$, and the potentials $\varphi$, $\phi$, reduce to two sets of equations when motions independent of the $x_3$ coordinate are considered. In the present paper the equations of interest are those governing the $u_3$ component of the displacement and the potentials $\varphi$, $\phi$, and can be written in the form

$$c_{44}\nabla^2 u_3 + e_{15}\nabla^2\varphi + q_{15}\nabla^2\phi = \rho\ddot{u}_3,$$

$$e_{15}\nabla^2 u_3 - \varepsilon_{11}\nabla^2\varphi - d_{11}\nabla^2\phi = 0, \qquad (1)$$

$$q_{15}\nabla^2 u_3 - d_{11}\nabla^2\varphi - \mu_{11}\nabla^2\phi = 0,$$

where $\nabla^2 = \partial^2/\partial x_1^2 + \partial^2/\partial x_2^2$ is the two-dimensional Laplacian operator, $\rho$ is the mass density, $c_{44}$, $e_{15}$, $\varepsilon_{11}$, $q_{15}$, $d_{11}$ and $\mu_{11}$ are elastic, piezoelectric, dielectric, piezomagnetic, electromagnetic and magnetic constants, and the superposed dot indicates differentiation with respect to time. The constitutive equations which relate the stresses $T_{ij}$ ($i,j=1,2,3$), the electric displacements $D_i$ ($i=1,2,3$) and the magnetic induction $B_i$ ($i=1,2,3$) to $u_3$, $\varphi$ and $\phi$ are

$$T_1 = T_2 = T_3 = T_{12} = 0, \; D_3 = 0, \; B_3 = 0,$$

$$T_{23} = c_{44}u_{3,2} + e_{15}\varphi_{,2} + q_{15}\phi_{,2}, \quad T_{13} = c_{44}u_{3,1} + e_{15}\varphi_{,1} + q_{15}\phi_{,1},$$

$$D_1 = e_{15}u_{3,1} - \varepsilon_{11}\varphi_{,1} - d_{11}\phi_{,1}, \quad D_2 = e_{15}u_{3,2} - \varepsilon_{11}\varphi_{,2} - d_{11}\phi_{,2}, \qquad (2)$$

$$B_1 = q_{15}u_{3,1} - d_{11}\varphi_{,1} - \mu_{11}\phi_{,1}, \quad B_2 = q_{15}u_{3,2} - d_{11}\varphi_{,2} - \mu_{11}\phi_{,2}.$$

Solving Eqs. (1) for $\nabla^2 u_3$, $\nabla^2\varphi$ and $\nabla^2\phi$ it has been found that after defining functions $\psi$ and $\chi$ by

$$\psi = \varphi - mu_3, \quad \chi = \phi - nu_3, \qquad (3)$$

the solution of Eqs. (1) is reduced to the solution of

$$\nabla^2 u_3 = \rho\tilde{c}_{44}^{-1}\ddot{u}_3, \quad \nabla^2\psi = 0, \quad \nabla^2\chi = 0, \qquad (4)$$

where

$$m = \frac{e_{15}\mu_{11} - q_{15}d_{11}}{\varepsilon_{11}\mu_{11} - d_{11}^2}, \quad n = \frac{q_{15}\varepsilon_{11} - e_{15}d_{11}}{\varepsilon_{11}\mu_{11} - d_{11}^2}, \qquad (5)$$

and

$$\tilde{c}_{44} = c_{44} + \left(e_{15}^2\mu_{11} - 2e_{15}q_{15}d_{11} + q_{15}^2\varepsilon_{11}\right)\big/\left(\varepsilon_{11}\mu_{11} - d_{11}^2\right)$$

$$= \bar{c}_{44}^e + \varepsilon_{11}^{-1}\left(d_{11}e_{15} - q_{15}\varepsilon_{11}\right)^2\big/\left(\varepsilon_{11}\mu_{11} - d_{11}^2\right)$$

$$= \bar{c}_{44}^m + \mu_{11}^{-1}(d_{11}q_{15} - e_{15}\mu_{11})^2 / (\varepsilon_{11}\mu_{11} - d_{11}^2). \tag{6}$$

In Eqs. (6) $\tilde{c}_{44}$ is magneto-electro-elastically stiffened elastic constant, $\bar{c}_{44}^e = c_{44} + e_{15}^2/\varepsilon_{11}$ is electro-elastically stiffened elastic constant and $\bar{c}_{44}^m = c_{44} + q_{15}^2/\mu_{11}$ is magneto-elastically stiffened elastic constant. With the analogy to the electro-mechanical coupling coefficient $k_e^2 = e_{15}^2/(\varepsilon_{11}\bar{c}_{44}^e)$ and the magneto-mechanical coupling coefficient $k_m^2 = q_{15}^2/(\mu_{11}\bar{c}_{44}^m)$ introduce the magneto-electro-mechanical coupling coefficient

$$k_{em}^2 = \tilde{c}_{44}^{-1}(e_{15}^2\mu_{11} - 2e_{15}q_{15}d_{11} + q_{15}^2\varepsilon_{11})/(\varepsilon_{11}\mu_{11} - d_{11}^2)$$

$$= e_{15}^2/(\tilde{c}_{44}\varepsilon_{11}) + \tilde{c}_{44}^{-1}\varepsilon_{11}^{-1}(q_{15}\varepsilon_{11} - e_{15}d_{11})^2/(\varepsilon_{11}\mu_{11} - d_{11}^2)$$

$$= q_{15}^2/(\tilde{c}_{44}\mu_{11}) + \tilde{c}_{44}^{-1}\mu_{11}^{-1}(e_{15}\mu_{11} - q_{15}d_{11})^2/(\varepsilon_{11}\mu_{11} - d_{11}^2). \tag{7}$$

From Eqs. (5)-(7) it follows that

$$e_{15}m + q_{15}n = \tilde{c}_{44}k_{em}^2, \quad \varepsilon_{11}m + d_{11}n = e_{15}, \quad d_{11}m + \mu_{11}n = q_{15},$$

$$(e_{15}\mu_{11} - q_{15}d_{11})m = \tilde{c}_{44}\mu_{11}k_{em}^2 - q_{15}^2, \tag{8}$$

$$(q_{15}\varepsilon_{11} - e_{15}d_{11})n = \tilde{c}_{44}\varepsilon_{11}k_{em}^2 - e_{15}^2.$$

Using the introduced functions $\psi$ and $\chi$ and the magneto-electro-elastically stiffened elastic constant, the constitutive Eqs. (2) can be written in the following form:

$$T_{23} = \tilde{c}_{44}u_{3,2} + e_{15}\psi_{,2} + q_{15}\chi_{,2}, \quad T_{13} = \tilde{c}_{44}u_{3,1} + e_{15}\psi_{,1} + q_{15}\chi_{,1},$$

$$D_1 = -\varepsilon_{11}\psi_{,1} - d_{11}\chi_{,1}, \quad D_2 = -\varepsilon_{11}\psi_{,2} - d_{11}\chi_{,2}, \tag{9}$$

$$B_1 = -d_{11}\psi_{,1} - \mu_{11}\chi_{,1}, \quad B_2 = -d_{11}\psi_{,2} - \mu_{11}\chi_{,2}.$$

From the condition of the positiveness of energy using Eqs. (2) one has that

$$c_{44} > 0, \quad \varepsilon_{11} > 0, \quad \mu_{11} > 0, \quad \varepsilon_{11}\mu_{11} - d_{11}^2 > 0. \tag{10}$$

From Eqs. (6), (7) and (10) one has that

$\tilde{c}_{44} \geq c_{44}$, $\tilde{c}_{44} \geq \bar{c}_{44}^e$, $\tilde{c}_{44} \geq \bar{c}_{44}^m$;

$\tilde{c}_{44} = c_{44}$ if and only if $e_{15} = 0$, $q_{15} = 0$;

$\tilde{c}_{44} = \bar{c}_{44}^e$ if and only if $d_{11}e_{15} = \varepsilon_{11}q_{15}$; $\tag{11}$

$\tilde{c}_{44} = \bar{c}_{44}^m$ if and only if $\mu_{11}e_{15} = d_{11}q_{15}$;

and

$$k_{em}^2 \geq e_{15}^2 / (\tilde{c}_{44}\varepsilon_{11}), \ k_{em}^2 \geq q_{15}^2 / (\tilde{c}_{44}\mu_{11}), \ 0 \leq k_{em} < 1;$$

$$k_{em}^2 = e_{15}^2 / (\tilde{c}_{44}\varepsilon_{11}) \text{ if and only if } d_{11}e_{15} = \varepsilon_{11}q_{15};$$

$$k_{em}^2 = q_{15}^2 / (\tilde{c}_{44}\mu_{11}) \text{ if and only if } \mu_{11}e_{15} = d_{11}q_{15}; \qquad (12)$$

$$k_{em} = 0 \text{ if and only if } e_{15} = 0, \ q_{15} = 0;$$

if $d_{11} = 0$ then $\tilde{c}_{44} = c_{44} + e_{15}^2 \varepsilon_{11}^{-1} + q_{15}^2 \mu_{11}^{-1}$, $k_{em}^2 = e_{15}^2 / (\tilde{c}_{44}\varepsilon_{11}) + q_{15}^2 / (\tilde{c}_{44}\mu_{11})$.

Introduce short notations $e = e_{15}$, $\mu = \mu_{11}$, $d = d_{11}$, $\varepsilon = \varepsilon_{11}$, $q = q_{15}$, $c = c_{44}$, $\bar{c}^e = \bar{c}_{44}^e$, $\bar{c}^m = \bar{c}_{44}^m$, $\tilde{c} = \tilde{c}_{44}$, $w = u_3$, $T = T_{23}$, $D = D_2$, $B = B_2$ and use subscripts $A$ and $B$ to refer to the half-spaces $x_2 > 0$ and $x_2 < 0$, respectively. Since the materials in the half-spaces $x_2 > 0$ and $x_2 < 0$ are identical, one has that $e_A = e_B = e$, $\mu_A = \mu_B = \mu$, $d_A = d_B = d$, $\varepsilon_A = \varepsilon_B = \varepsilon$, $q_A = q_B = q$, $c_A = c_B = c$, $\bar{c}_A^e = \bar{c}_B^e = \bar{c}^e$, $\bar{c}_A^m = \bar{c}_B^m = \bar{c}^m$, $\tilde{c}_A = \tilde{c}_B = \tilde{c}$.

The conditions at infinity require that

$$w_A, \ \varphi_A, \ \phi_A \to 0 \text{ as } x_2 \to \infty,$$

$$w_B, \ \varphi_B, \ \phi_B \to 0 \text{ as } x_2 \to -\infty, \qquad (13)$$

and the mechanical boundary conditions on the plane boundaries of the half-spaces require that

$$T_A = T_B = 0 \text{ on } x_2 = 0. \qquad (14)$$

Consider the possibility of a solution of Eqs. (3)-(4) of the form

$$w_A = w_{0A} \exp(-\xi_2 x_2) \exp[i(\xi_1 x_1 - \omega t)],$$

$$\psi_A = \psi_{0A} \exp(-\xi_1 x_2) \exp[i(\xi_1 x_1 - \omega t)], \qquad (15)$$

$$\chi_A = \chi_{0A} \exp(-\xi_1 x_2) \exp[i(\xi_1 x_1 - \omega t)],$$

in the half-space $x_2 > 0$ and of the form

$$w_B = w_{0B} \exp(\xi_2 x_2) \exp[i(\xi_1 x_1 - \omega t)],$$

$$\psi_B = \psi_{0B} \exp(\xi_1 x_2) \exp[i(\xi_1 x_1 - \omega t)], \qquad (16)$$

$$\chi_B = \chi_{0B} \exp(\xi_1 x_2) \exp[i(\xi_1 x_1 - \omega t)],$$

in the half-space $x_2 < 0$. These expressions satisfy the conditions (13) if $\xi_1 > 0$ and $\xi_2 > 0$; the second and the third of Eqs. (4) are identically satisfied and the first of Eqs. (4) requires

$$\tilde{c}\left(\xi_1^2 - \xi_2^2\right) = \rho\omega^2. \tag{17}$$

Now the mechanical boundary conditions (14) together with different magneto-electrical contact conditions on $x_2 = 0$ must be satisfied. In the present paper the following cases of the magneto-electrical contact conditions on $x_2 = 0$ are of our interest:

1a) $D_A = D_B = 0$, $B_A = B_B$, $\phi_A = \phi_B$;

1b) $D_A = D_B = 0$, $\phi_A = \phi_B = 0$;

1c) $D_A = D_B = 0$, $B_A = 0$, $\phi_B = 0$;

2a) $\varphi_A = \varphi_B = 0$, $B_A = B_B = 0$;

2b) $D_A = D_B$, $\varphi_A = \varphi_B$, $B_A = B_B = 0$;

2c) $\varphi_A = 0$, $D_B = 0$, $B_A = B_B = 0$;

3a) $D_A = D_B$, $\varphi_A = \varphi_B$, $B_A = B_B$, $\phi_A = \phi_B$;

3b) $\varphi_A = \varphi_B = 0$, $\phi_A = \phi_B = 0$;

3c) $D_A = 0$, $\varphi_B = 0$, $B_A = 0$, $\phi_B = 0$; \hfill (18)

4a) $\varphi_A = \varphi_B = 0$, $B_A = B_B$, $\phi_A = \phi_B$;

4b) $\varphi_A = \varphi_B = 0$, $B_A = 0$, $\phi_B = 0$;

5a) $D_A = D_B$, $\varphi_A = \varphi_B$, $\phi_A = \phi_B = 0$;

5b) $\varphi_A = 0$, $D_B = 0$, $\phi_A = \phi_B = 0$;

6) $\varphi_A = 0$, $D_B = 0$, $B_A = 0$, $\phi_B = 0$;

7a) $\varphi_A = 0$, $D_B = 0$, $B_A = B_B$, $\phi_A = \phi_B$;

7b) $D_A = D_B$, $\varphi_A = \varphi_B$, $B_A = 0$, $\phi_B = 0$;

8) $D_A = D_B = 0$, $B_A = B_B = 0$.

Each of the 17 groups of conditions in Eqs. (18) together with Eqs. (14), (15)-(16) leads to a system of six linear homogeneous algebraic equations for $w_{0A}$, $\psi_{0A}$, $\chi_{0A}$, $w_{0B}$, $\psi_{0B}$, $\chi_{0B}$, the existence of nonzero solution of which requires that the determinant of that system be equal to zero. This condition for the determinant together with Eq. (17) determines the surface wave velocities $V_s = \omega/\xi_1$. In the case of 1a) of Eqs. (18) this procedure leads to a surface wave with the following velocity:

$$V_{s1}^2 = (\tilde{c}/\rho)\left(1 - \left[k_{em}^2 - e^2/(\tilde{c}\varepsilon)\right]^2\right). \tag{19}$$

The same velocity is obtained in the cases 1b) and 1c). Each of the cases 2a), 2b) and 2c) leads to a surface wave with velocity

$$V_{s2}^2 = (\tilde{c}/\rho)\left(1 - \left[k_{em}^2 - q^2/(\tilde{c}\mu)\right]^2\right), \tag{20}$$

and each of the cases 3a), 3b) and 3c) leads to a surface wave with velocity

$$V_{s3}^2 = (\tilde{c}/\rho)\left(1 - k_{em}^4\right). \tag{21}$$

The case 8) does not lead to any surface wave. Each of the cases from 4a) to 7b) leads to its own pair of surface waves with velocities

$V_{s2}$, $V_{s3}$ in the cases 4a) and 4b);

$V_{s1}$, $V_{s3}$ in the cases 5a) and 5b);

$V_{s1}$, $V_{s2}$ in the case 6);

$V_{s4}$, $V_{s5}$ in cases 7a) and 7b);

where

$$V_{s4}^2 = (\tilde{c}/\rho)(1-\alpha^2), \quad V_{s5}^2 = (\tilde{c}/\rho)(1-\beta^2),$$

$$\alpha = \frac{1}{2}\frac{\varepsilon\mu}{2\varepsilon\mu - d^2}\left(\frac{3\varepsilon\mu - d^2}{\varepsilon\mu}k_{em}^2 - \frac{e^2}{\tilde{c}\varepsilon} - \frac{q^2}{\tilde{c}\mu} - \sqrt{Q}\right),$$

$$\beta = \frac{1}{2}\frac{\varepsilon\mu}{2\varepsilon\mu - d^2}\left(\frac{3\varepsilon\mu - d^2}{\varepsilon\mu}k_{em}^2 - \frac{e^2}{\tilde{c}\varepsilon} - \frac{q^2}{\tilde{c}\mu} + \sqrt{Q}\right), \tag{22}$$

$$Q = \left(\frac{3\varepsilon\mu - d^2}{\varepsilon\mu}k_{em}^2 - \frac{e^2}{\tilde{c}\varepsilon} - \frac{q^2}{\tilde{c}\mu}\right)^2 - 4\frac{2\varepsilon\mu - d^2}{\varepsilon\mu}\left(k_{em}^2 - \frac{q^2}{\tilde{c}\mu}\right)\left(k_{em}^2 - \frac{e^2}{\tilde{c}\varepsilon}\right)$$

$$= \frac{\left(e^2\mu - 2eqd + q^2\varepsilon\right)^2 + 2(q\varepsilon - ed)^2 q^2 + 2(e\mu - qd)^2 e^2 + \left(\mu e^2 - \varepsilon q^2\right)^2}{(\varepsilon\mu\tilde{c})^2}.$$

From Eqs. (22) it follows that

$$(1-\beta)(1-\alpha) = \frac{\varepsilon\mu}{2\varepsilon\mu - d^2}\left[\left(\frac{\varepsilon\mu - d^2}{\varepsilon\mu} + (1 - k_{em}^2) + \frac{e^2}{\tilde{c}\varepsilon} + \frac{q^2}{\tilde{c}\mu}\right)(1 - k_{em}^2) + \frac{e^2 q^2}{\tilde{c}^2\varepsilon\mu}\right]. \tag{23}$$

Using Eqs. (10)-(12), (22) and (23) one has that

$$Q \geq 0; \quad 0 \leq \alpha \leq \beta < 1; \tag{24}$$

$\alpha = 0$ if and only if $ed = \varepsilon q$ or $\mu e = qd$.

From Eqs. (10)-(12), (22) and (23) it also follows that each of the equalities $Q = 0$, $\beta = 0$, $\alpha = \beta$ takes place if and only if the magneto-electro-mechanical coupling coefficient is equal to zero.

If the magneto-electro-elastic material degenerates to a piezoelectric material, so that $q \to 0$, $d \to 0$, the surface waves that have velocities $V_{s1}$, $V_{s4}$ disappear and

$$V_{s2}, V_{s3}, V_{s5} \to V_{bg} = \sqrt{\left(\overline{c}^e/\rho\right)\left(1 - k_e^4\right)}, \tag{25}$$

so that the possibility for surface waves to be guided in pairs disappears, and the number of different surface wave velocities decreases from 5 to 1.

**REFERENCES**

1 J.L. Bleustein: Appl. Phys. Lett. **13**, 412 (1968)

2 Y.V. Gulyaev: JETP Lett. **9**, 37 (1968)

3 E. Danicki: Appl. Phys. Lett. **64**, 969 (1994)

4 C.W. Nan: Phys. Rev. B **50**, 6082 (1994)

5 K. Srinivas, G. Prasad, T. Bhimasankaram and S. V. Suryanarayana: Modern Phys. Lett. B **14**, 663 (2000)

6 K. Mori and M. Wuttig: Appl. Phys. Lett. **81**, 100 (2002)

7 P. Yang, K. Zhao, Y. Yin, J.G. Wan, and J.S. Zhu: Appl. Phys. Lett. **88**, 172903 (2006)

8 S. Srinivas, J.Y. Li, Y.C. Zhou, A.K. Soh: J. Appl. Phys. **99**, 043905 (2006)